# BERTopic for Topic Modeling of Hindi Short Texts: A Comparative Study

**Atharva Mutsaddi, Anvi Jamkhande, Aryan Thakre, Yashodhara Haribhakta**

Department of Computer Science and Engineering, COEP Technological University

`{atharvaam21, jamkhandeaa21, aryanst21, ybl}.comp@coeptech.ac.in`

## Abstract

As short text data in native languages like Hindi increasingly appear in modern media, robust methods for topic modeling on such data have gained importance. This study investigates the performance of BERTopic in modeling Hindi short texts, an area that has been under-explored in existing research. Using contextual embeddings, BERTopic can capture semantic relationships in data, making it potentially more effective than traditional models, especially for short and diverse texts. We evaluate BERTopic using 6 different document embedding models and compare its performance against 8 established topic modeling techniques, such as Latent Dirichlet Allocation (LDA), Non-negative Matrix Factorization (NMF), Latent Semantic Indexing (LSI), Additive Regularization of Topic Models (ARTM), Probabilistic Latent Semantic Analysis (PLSA), Embedded Topic Model (ETM), Combined Topic Model (CTM), and Top2Vec. The models are assessed using coherence scores across a range of topic counts. Our results reveal that BERTopic consistently outperforms other models in capturing coherent topics from short Hindi texts.

## 1 Introduction

Topic modeling is a widely-used technique in text mining that identifies underlying themes within textual data. BERTopic, a newer model in this field, has demonstrated its effectiveness by using pre-trained document embedding models and unsupervised clustering algorithms to form topic groups with high semantic coherence (Grootendorst, 2022). Unlike traditional models, BERTopic's use of embeddings allows it to capture contextual information, such as identifying named entities and associating them with relevant topic clusters that older approaches struggle with (Peters et al., 2018; Liu et al., 2019). Existing research on topic modeling for Hindi texts has largely focused on traditional methods that rely on probabilistic frameworks and matrix factorisation, which often overlook natural language semantics (Ray et al., 2019). Also, these studies primarily focus on long text documents, leaving a gap in the exploration of short Hindi texts, which are increasingly common in todays digital landscape.

Topic modeling in Hindi faces several unique challenges. Hindi does not use capitalisation to differentiate proper nouns from other word forms, complicating named entity recognition. Additionally, the lack of standardised spelling leads to multiple variations of the same word (Figure 1), creating ambiguity. Hindi also often employs repetitive expressions for emphasis, which can affect tokenization and cross-language natural language processing tasks (Ray et al., 2019).

This study aims to demonstrate that traditional topic models often fall short in capturing the semantic meaning of Hindi text due to these inherent challenges and struggle with the nuances of short texts where semantic meaning is more compressed and context-dependent. The contributions of this paper are as follows:

- Conducting a comprehensive comparison of BERTopic's performance across several aspects:
  - Evaluating BERTopic using different sentence transformer models such as `HindSBERT-STS` (Joshi et al., 2022), `XLM-R` (XLM-RoBERTa) (Conneau et al., 2020), `IndicBERT` (Kakwani et al., 2020), and `mBERT` (Multilingual BERT) (Devlin et al., 2018), and analysing results using coherence metrics $c_v$ (coherence value) (Röder et al., 2015) and $c_{\text{NPMI}}$ (Normalised Pointwise Mutual Information) (Bouma, 2009) to identify the optimal transformer model.
  - Comparing BERTopic with traditional

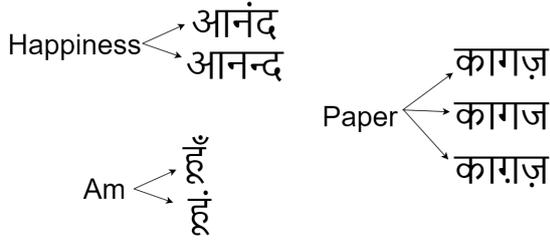

Figure 1: Multiple ways of spelling "Happiness","Paper" and "Am" in Hindi

topic modeling methods like LDA, NMF, LSI, ARTM, and PLSA, to show that BERTopic consistently outperforms these models.

– Exploring additional transformer-based models such as Top2vec (Angelov, 2020), Embedded Topic Model (ETM) (Dieng et al., 2020) and Combined Topic Model (CTM) (Terragni et al., 2021) to evaluate their relative performance.

- Demonstrating BERTopics effectiveness in addressing the challenges of modeling short Hindi texts by handling compressed and context-dependent semantics, as evidenced by the comparative analysis.

- Contributing to the study of under-explored languages such as Hindi, by highlighting the benefits of advanced topic models for enhancing semantic coherence and topic extraction in non-English languages.

## 2 Previous Work

Topic modeling studies on Hindi text have predominantly relied on traditional frameworks, such as probabilistic models and matrix factorisation techniques, while newer approaches remain largely under-explored. Furthermore, most of these studies have focused on long text documents, leaving the effectiveness of topic modeling techniques on short texts inadequately examined.

Ray et al. (2019) provided an overview of various topic modeling approaches for Hindi text, including methods like NMF (Lee and Seung, 1999), LSI (Deerwester et al., 1990), and LDA (Jelodar et al., 2019), as well as tools and Java packages used in these models. However, their work predates the development of BERTopic and does not address its application. Similarly, Srivastav and Singh (2022) investigated the use of models such as LDA, Doc2Vec, and Hierarchical Dirichlet Process (HDP) (Teh et al., 2006) for identifying the main topics in news articles in both Hindi and English. Their study, however, also did not consider newer topic modeling techniques like BERTopic. Panigrahi et al. (2018) explored an embedding-based clustering approach, using Word2Vec (Mikolov, 2013) to generate word embeddings from a corpus of Hindi Wikipedia articles, which were subsequently clustered into topic groups. While this study adopted an approach similar to BERTopic, it did not specifically focus on short texts or use more advanced document embedding models.

While BERTopic's effectiveness on Hindi texts remains unexplored, some studies have evaluated its performance in other non-English languages. Abuzayed and Al-Khalifa (2021) compared BERTopic using different sentence transformers against LDA and NMF on Arabic news articles, and found that BERTopic formed more coherent topic clusters by evaluating them using Normalised Pointwise Mutual Information (Bouma, 2009). Another study (Abdelrazek et al., 2022) compared the computational cost and topic quality of LDA, ETM, CTM, NMF, and two BERTopic variants on Arabic data, concluding that BERTopic outperformed other models in coherence scores. Although these studies focused on longer texts, Medvecki et al. (2024) demonstrated that BERTopic produced more informative clusters than LDA and NMF when applied to Serbian tweets, showing its efficiency in modeling short text data in other languages.

Although BERTopic's superior performance has been proven for some other non-English languages, its effectiveness for Hindi, especially on short texts, remains un-examined. Given Hindi's unique linguistic challenges (Ray et al., 2019) and its status as the third most spoken language globally, it is important to evaluate BERTopic's performance, particularly for Hindi short texts, which are increasingly common in modern media.

## 3 Methodology

This section gives an overview about the dataset we used, the topic models we evaluated and the method of evaluation used for this comparison.

## 3.1 Dataset

We used the IIT Patna Reviews dataset, a well-regarded dataset used for evaluating and training Hindi natural language processing models for the task of sentiment analysis. This dataset contains short text product reviews written in the Devanagari script, each mapped to its corresponding sentiment. For this study, we focused on modeling the reviews and did not use the sentiment mappings.

For pre-processing, we used the Hindi language stop words list compiled by Jha et al. (2018), to identify and remove stop words. We also removed punctuation marks, URLs, username references, extra spaces, hashtags, and leading and trailing quotations to reduce noise in the data.

## 3.2 Evaluation

For the performance evaluation of these topic models, we used coherence value ($c_v$) and Normalised Pointwise Mutual Information ($c_{\text{NPMI}}$) to assess the quality of topics formed (Röder et al., 2015; Bouma, 2009). The $c_v$ metric evaluates the semantic coherence of a set of words which represent a topic using word co-occurrence graphs. The $c_{\text{NPMI}}$ metric evaluates the word associations within a topic cluster, assessing how strongly the words are related.

For each model, the average $c_v$ and $c_{\text{NPMI}}$ scores were calculated across the topic clusters, and these scores were used for comparison. We perform these calculations for topic counts ranging from 5 to 210. This range was chosen because BERTopic scores generally stabilise beyond 210 topics, indicating that adding more topics does not significantly alter topic coherence. Also, considering the overall size of our dataset, 210 topic clusters were deemed sufficient for meaningful topic labeling. While BERTopic can automatically determine the optimal number of topic clusters to form, we specified the number of clusters in this comparison to ensure a consistent basis for evaluating its performance against other models, which require a predefined number of clusters.

The interpretation of coherence scores in topic modeling is subject to debate. Previous studies (He et al., 2009, 2008; Newman et al., 2011; Das Dawn et al., 2024) suggest that a lower $c_v$ score, typically below 0.4, indicates overly generalised topic clusters, while scores above 0.7 might suggest more specialised ones. Despite this debate, there is consensus that higher $c_{\text{NPMI}}$ scores reflect stronger word relations within the topic clusters (Abuzayed and Al-Khalifa, 2021; Medvecki et al., 2024).

## 3.3 BERTopic

BERTopic uses embedding models to understand the semantic meaning and context in which words are used (Grootendorst, 2022), making it well-suited for modeling Hindi short texts. It employs a dimensionality reduction algorithm like UMAP (Uniform Manifold Approximation and Projection) (McInnes et al., 2018), followed by an unsupervised clustering algorithm like HDBSCAN (Hierarchical Density-Based Spatial Clustering of Applications with Noise) (McInnes et al., 2017) or KMeans (MacQueen et al., 1967) to create coherent topic clusters. Our experiment involved using various sentence transformers with BERTopic and comparing their relative performance to choose the most optimal one for further comparisons.

The transformers we compared were:

- `XLM-R` (xlm-roberta-base) (Conneau et al., 2020)

- `IndicBERT`, which is a transformer fine-tuned for Indic languages (Kakwani et al., 2020).

- `HindSBERT-STS`, a transformer designed for semantic textual similarity tasks in Hindi (Joshi et al., 2022), using SBERT (Sentence-BERT) (Reimers and Gurevych, 2019).

- `mBERT` (bert-base-multilingual-cased for `mBERT-Cased` and bert-base-multilingual-uncased for `mBERT-Uncased`) (Devlin et al., 2018)

These embedding models were selected based on their ability to capture the semantic and contextual meaning of Hindi, which is essential for modeling short text reviews. Comparing language-specific and multilingual embedding models is vital for this analysis. Language-specific models, such as `HindSBERT-STS` and `IndicBERT`, are trained predominantly on Hindi corpora and are well-suited to handle features unique to Hindi, including compound verbs, spelling variations, and idiomatic expressions. On the other hand, multilingual embeddings are trained on a broader and more diverse set of languages, enabling them to leverage cross-lingual transfer for improved performance on low-resource languages by identifying shared linguistic patterns. Additionally, multilingual embed-dings often exhibit greater robustness in tasks like named entity recognition and cross-lingual reference handling, making them particularly advantageous for processing multilingual or code-mixed content. This comparison highlights the distinct strengths of each approach, providing valuable insights for selecting embeddings based on specific use cases.

### 3.4 Comparative Models

We compared BERTopic with the following models:

- **LDA-Based Models**: These models utilise the LDA framework to identify topic distributions within the text. We compared the following variants:
    - LDA (Latent Dirichlet Allocation) (Blei et al., 2003).
    - ARTM (Additive Regularisation of Topic Models) (Vorontsov and Potapenko, 2015).
    - ETM (Embedded Topic Model) (Dieng et al., 2020) with the same sentence transformers as discussed in subsection 3.3. We consider the best transformer for further comparison with the other LDA based approaches.

    We consider the best variant of LDA for further comparison with other models.

- **Other Topic Modeling Approaches**:
    - NMF (Non-negative Matrix Factorisation) (Lee and Seung, 1999).
    - PLSA (Probabilistic Latent Semantic Analysis) (Hofmann, 1999).
    - LSI (Latent Semantic Indexing) (Deerwester et al., 1990).
    - CTM (Combined Topic Model). Specifically the `Octis` implementation of it (Terragni et al., 2021).
    - Top2Vec (Angelov, 2020). Since we cannot specify the number of topic clusters in Top2Vec, we compared the best scores it achieved with four different embedding models, namely- distiluse-base-multilingual-cased, universal-sentence-encoder, universal-sentence-encoder-multilingual and doc2vec. We later considered the best embedding model with Top2Vec for further comparison.

### 3.5 Implementation Details

All experiments were conducted using Google Colaboratory with the following Python tools:

- `Sklearn` for implementing ETM and NMF.
- `Gensim` for LSI and LDA Multicore.
- `Bigartm` for PLSA and ARTM.
- `Octis` library for CTM.
- `sentence-transformers` for Hugging Face models: `xlm-roberta-base`, `indic-bert`, `HindSBERT-STS`, `bert-base-multilingual-cased`, and `bert-base-multilingual-uncased`.

## 4 Results and Analysis

We compared 20 models, each utilising different approaches, including embedding-based models like BERTopic and ETM with the 5 embedding models mentioned in subsection 3.3, hybrid models like CTM and Top2Vec using 4 pre-trained embedding models (subsection 3.4), probabilistic models like LDA, ARTM, and PLSA, and matrix factorisation models such as NMF and LSI. Following are our findings:

### 4.1 Comparison of LDA-Based Models

After evaluating multiple LDA variants, we found that ETM with `HindSBERT-STS` yielded the most coherent topics, outperforming the other embedding models for majority of topic counts (Figures 2 , 3). Specifically, ETM achieved a $c_v$ and $c_{\text{NPMI}}$ score of 0.71 and 0.089 respectively for 205 topics.

The $c_v$ scores for ETM model using `HindSBERT-STS` suggest a balance in topic specificity, avoiding both highly specific and overly generalised clusters (He et al., 2009, 2008; Newman et al., 2011; Das Dawn et al., 2024).

ARTM performed better than all LDA variants in terms of $c_v$ scores, particularly in the 5 to 20 topic range, but it failed to maintain this trend for higher topic counts (Figures 4 , 5).

The performance of the traditional LDA model declined as the number of topics increased, showing its limitations in handling short text data with fewer words available for topic extraction (Qiang et al., 2022; Aggarwal and Zhai, 2012).

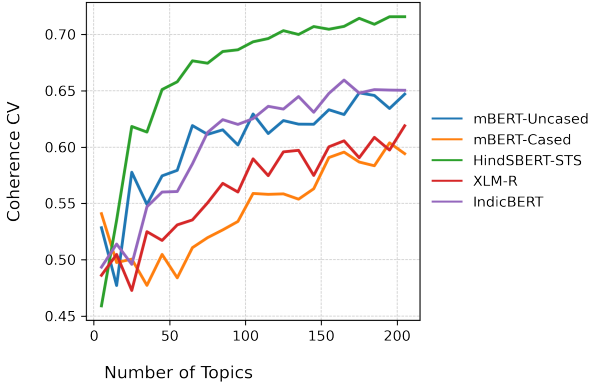

Figure 2: $c_v$ scores of ETM with different sentence transformers

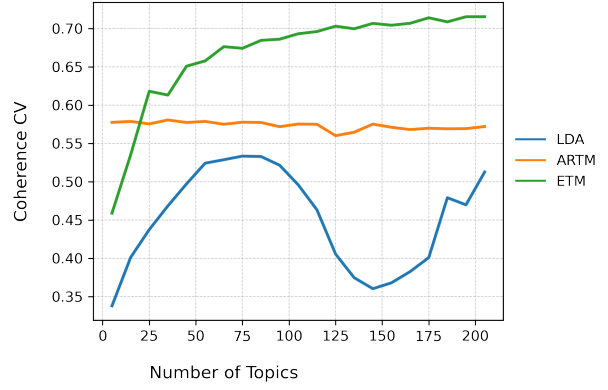

Figure 4: Comparison of $c_v$ Scores for LDA-Based Models

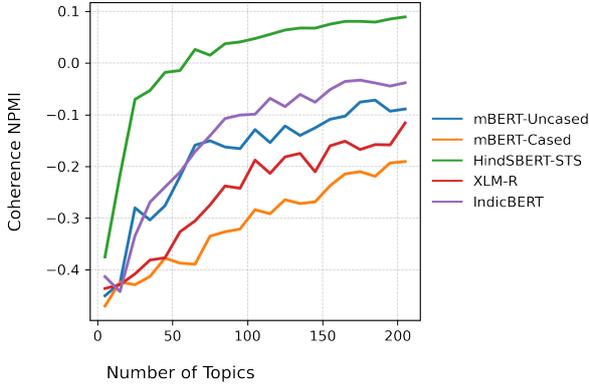

Figure 3: $c_{\text{NPMI}}$ scores of ETM with different sentence transformers

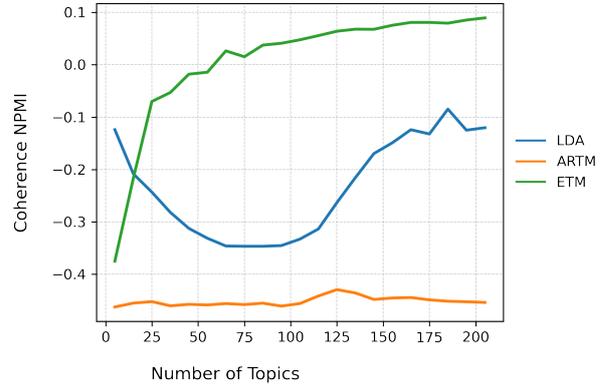

Figure 5: Comparison of NPMI Scores for LDA-Based Models

Hence we can see that ETM is the best LDA variant amongst the ones we have evaluated.

### 4.2 Evaluation of Embedding Models with BERTopic

Figures 6 and 7 show that `mBERT-Uncased` consistently provides better results than the other sentence transformers when used with BERTopic. The high $c_v$ scores achieved by `mBERT-Uncased` and `XLM-R` at larger topic counts suggest the formation of dense, specialised clusters (He et al., 2009, 2008; Newman et al., 2011) with strong semantic relationships among the words within these topics (Hadiat, 2022). Although `XLM-R` barely outperforms `mBERT-Uncased` in $c_v$ scores from 170 topics onward, its $c_{\text{NPMI}}$ scores are significantly worse across the entire topic count range (Figure 7).

While `mBERT-Cased` performs better than the other models at lower topic counts, its scores decreased significantly as the number of topics increased, leading to its exclusion from further evaluation.

Additionally, due to poor performance, `HindSBERT-STS` and `IndicBERT` were not considered for further analysis.

### 4.3 Performance Analysis: Best BERTopic vs. Best LDA vs. Other Models

We found that BERTopic with the `mBERT-Uncased` embedding model outperformed other topic models for the majority of topic counts (Figures 8, 9). Table 1 presents the best coherence scores obtained by each model, along with the corresponding number of topics at which these scores were achieved.

BERTopic produced significantly higher coherence scores than all other models, with its $c_v$ scores being, on average, 19.8% higher than those of ETM with `HindSBERT-STS`, which ranked second (Figure 8). While ETM formed topic clusters with slightly higher $c_{\text{NPMI}}$ scores than BERTopic for 125 topics onwards, BERTopic showed better scores across most topic ranges, indicating more consistent performance.

NMF and PLSA demonstrated nearly identi-cal coherence scores, both ranking approximately third in $c_v$ scores and last in $c_{\text{NPMI}}$.

| Model | $c_{\text{NPMI}}$ | $c_v$ | Topic Count |
|---|---|---|---|
| **BERTopic [mBERT-Uncased]** | **0.07** | **0.76** | **95** |
| ETM [HindSBERT-STS] | 0.089 | 0.71 | 205 |
| Top2Vec [DBMC] | −0.48 | 0.54 | 45 |
| PLSA | −0.46 | 0.57 | 45 |
| ARTM | −0.42 | 0.56 | 125 |
| NMF | −0.44 | 0.56 | 35 |
| CTM | −0.38 | 0.48 | 135 |
| LDA | −0.12 | 0.38 | 165 |
| LSI | −0.08 | 0.30 | 15 |

Table 1: Best scores achieved by topic models on Hindi short text dataset

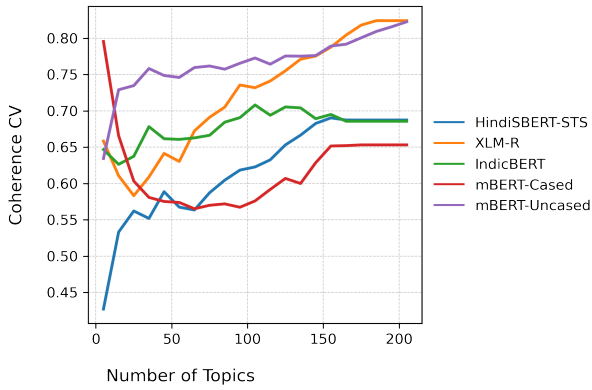

Figure 6: $c_v$ scores for BERTopic with Different Embedding Models

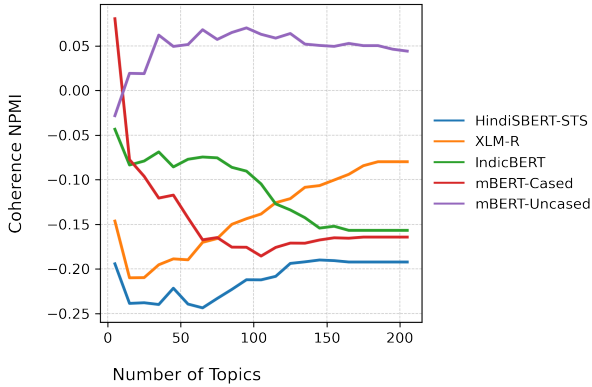

Figure 7: $c_{\text{NPMI}}$ scores of BERTopic with Different Embedding Models

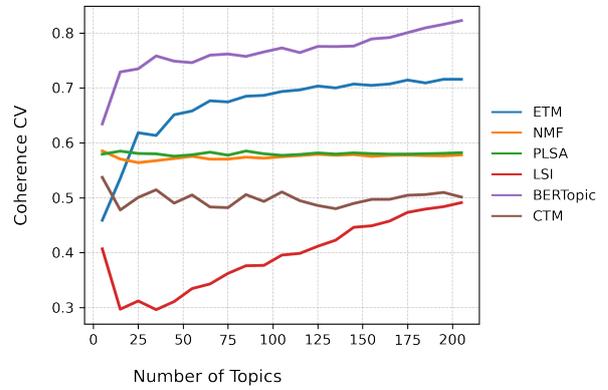

Figure 8: $c_v$ scores of BERTopic, ETM, NMF, PLSA, LSI and CTM

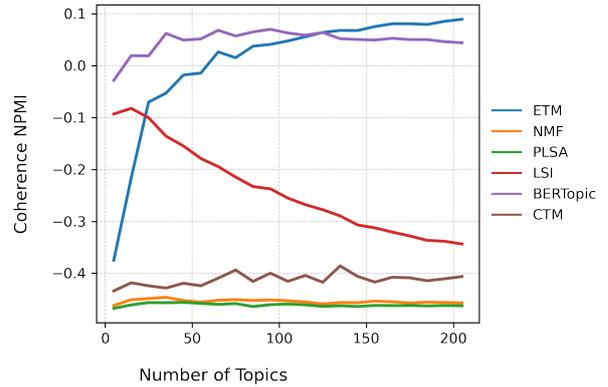

Figure 9: $c_{\text{NPMI}}$ scores of BERTopic, ETM, NMF, PLSA, LSI and CTM

For LSI, while the $c_v$ scores of its topic clusters increased for larger topic counts, the $c_{\text{NPMI}}$ scores declined over the same range. This suggests that as the topic count increased, LSI formed more specialised clusters with high word co-occurrence coherence. However, these clusters did not reflect strong word relations, as indicated by the decreasing $c_{\text{NPMI}}$ scores (Bouma, 2009; Röder et al., 2015).

As mentioned previously (subsection 3.4), we cannot specify the number of output clusters for Top2Vec, as it determines the optimal number of clusters autonomously (Angelov, 2020). Table 2 presents the best scores achieved by Top2Vec across various embedding models. We found that the distiluse-base-multilingual-cased model had the best relative performance, with a $c_{\text{NPMI}}$ score of -0.48 and a $c_v$ score of 0.54, for 45 topics. The universal-sentence-encoder-multilingual and universal-sentence-encoder models achieved their peak scores at 2 and 7 topics, respectively, suggesting that these models produced overly generalised topic clusters. This indicates that these embedding models were not robust enough to capture the diversity of topics present in Hindi short texts. Additionally, the low $c_v$ score for the topic clusters generated by doc2vec highlights a general lack of semantic coherence in the clusters formed.

| Embedding Model | $c_{\text{NPMI}}$ | $c_v$ | Topics |
|---|---|---|---|
| **distiluse-base-multilingual-cased** | **-0.48** | **0.54** | **45** |
| universal-sentence-encoder-multilingual | −0.45 | 0.53 | 2 |
| universal-sentence-encoder | −0.44 | 0.51 | 7 |
| doc2vec | −0.33 | 0.26 | 24 |

Table 2: Best scores achieved by Top2Vec using different embedding models

Overall, NMF, PLSA, LSI, CTM and Top2Vec, all had negative $c_{\text{NPMI}}$ scores, demonstrating poor performance in modeling Hindi short texts.

### 4.4 Qualitative Analysis of BERTopic Clusters

Apart from using $c_v$ and $c_{\text{NPMI}}$, we also qualitatively analysed the topic clusters formed by BERTopic through human evaluation, and verified the semantic coherence and relevance of the groupings. The dataset used for topic modeling encompassed a diverse range of topics, including film, tourism, and technology. For example, Figure 10 displays a word cloud for a topic cluster generated using `mBERT-Uncased` embeddings. As we can see, BERTopic successfully grouped reviews related to film and entertainment, capturing key terms such as फ़िल्म ("film") and किरदार ("character") in Hindi, reflecting its ability to form semantically coherent topic groups.

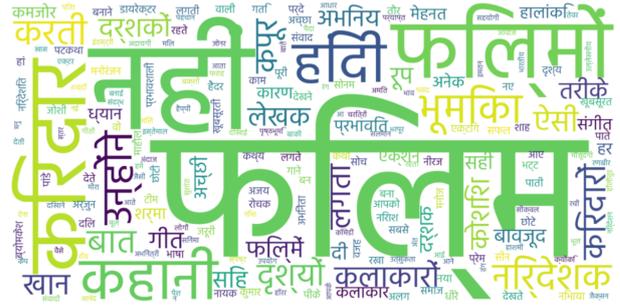

Figure 10: Word Cloud of a topic cluster formed by BERTopic

बात करें तो **एवेंजर्स: एज ऑफ अल्ट्रॉन** देखने को मिलती है।
**हैदर** में खोज करते हुए **नरेंद्र झा** की भूमिका है।
**जेम्स वान सॉ** के डायरेक्टर हैं।
हालिया रिलीज फिल्म **सैन एंड्रियास** एक ऐसी फिल्म है।
**वेलकम 2 कराची** एक नमूना है।

Figure 11: Some reviews belonging to the same cluster

If we examine a few reviews from this cluster (Figure 11), we see that BERTopic recognised the names of famous movies, such as एवेंजर्स: एज ऑफ अल्ट्रॉन (Avengers: Age of Ultron), हैदर (Haider), वेलकम 2 कराची (Welcome 2 Karachi), सॉ (Saw) and सैन एंड्रियास (San Andreas). It also identified the names of actors and directors, like नरेंद्र झा (Narendra Jha) and जेम्स वान (James Wan). BERTopic grouped these reviews into the same cluster, even though some had different word compositions. This indicates that the model effectively captured the contextual use of words, including named entities, with the help of advanced sentence transformers to form meaningful clusters. In contrast, traditional topic models which rely primarily on word frequency and co-occurrence, often fail to capture such semantic relationships, particularly in short texts.

## 5 Discussion

This study demonstrates that topic models utilising advanced sentence transformers, such as BERTopic and ETM, significantly outperform traditional models when modeling short texts. The success of these models can be attributed to their ability to capture semantic meaning beyond simple word co-occurrence patterns.

Traditional topic modeling algorithms like PLSA and LDA are widely used to uncover latent semantic structures in text corpora by relying on word co-occurrence patterns at the doc-ument level. However, these methods require a high frequency of word co-occurrences to generate meaningful topics, leading to significant performance degradation when applied to short texts where such information is sparse (Yin and Wang, 2014; Yan et al., 2013). Similarly, the performance of LSI declines over short texts as the detected topics become ambiguous, resulting in negative values in its decomposed matrices that are difficult to interpret (Murshed et al., 2023; Alghamdi and Alfalqi, 2015). Since many of these traditional models depend heavily on word frequency and co-occurrence, they are more sensitive to variations in spelling, a common issue in Hindi due to the lack of standardised spelling conventions (Ray et al., 2019). These limitations collectively undermine the reliability of traditional models in generating coherent topics from short text corpora.

## 6 Conclusion

We evaluated the performance of BERTopic relative to other topic models using coherence values ($c_v$) and normalised pointwise mutual information ($c_{\text{NPMI}}$) across a range of 5 to 210 topics. The results showed that BERTopic, particularly when used with `mBERT-uncased`, outperformed other models for the majority of topic counts. The ETM model, using `HindSBERT-STS`, ranked second, with better $c_{\text{NPMI}}$ scores than BERTopic beyond 125 topics, but consistently lower $c_v$ scores. Traditional topic models demonstrated poor performance, having negative $c_{\text{NPMI}}$ scores for the entire topic count range.

Qualitative analysis of BERTopic clusters revealed that it effectively grouped semantically similar reviews and accurately recognised named entities, a task at which traditional models struggle. The strong performance of both ETM and BERTopic suggests that leveraging advanced sentence transformers enhances the formation of coherent topic clusters.

We conclude that BERTopic is a promising approach for topic modeling on Hindi short text corpora, particularly when using multilingual transformers fine-tuned on Hindi. Its use can produce semantically coherent topic groups and better handle the unique linguistic complexities of the language. Potential applications include trend analysis, extracting business insights, analysing customer reviews and social media comments.

## 7 Future Work

Future work can explore the extent to which BERTopic results can be generalised to other Indo-Aryan languages, such as Sanskrit, Prakrit, Marathi, Konkani, and Nepali. These languages share linguistic similarities, including grammatical structure, Subject-Object-Verb (SOV) sentence ordering, and their use of the Devanagari script. This exploration would depend on the availability of sentence transformer models trained specifically for these languages.

Additionally, investigating the adaptability of BERTopic to other morphologically rich and low-resource languages, such as Tamil or Punjabi, could provide valuable insights into its broader applicability. Another promising direction is applying this approach to multilingual datasets or those containing code-mixed content, which reflects the increasing prevalence of mixed-language communication in digital spaces.

It would also be interesting to study how well BERTopic performs on longer texts compared to shorter ones for Indo-Aryan languages like Hindi, as evaluating BERTopic's ability to handle such texts could provide deeper insights into its capacity to model topics in languages with complex linguistic structures and ensure its effectiveness for use cases such as document-level topic extraction.

## Limitations

While this comparative study demonstrates the efficiency of BERTopic for topic modeling of Hindi short text reviews, there are some limitations to consider.

First, the IIT Patna Reviews Dataset, although a reputable and commonly used Hindi short text dataset for NLP research in Indian languages, is limited in size, containing only 5,225 reviews. Larger and domain-specific datasets could provide further insights into model performance and robustness. Due to the current lack of available benchmark datasets for Hindi short texts, we relied on this dataset for our study.

The dataset may also exhibit biases that influence the results. For instance, a representation bias exists, with a higher concentration of reviews on popular topics like movies and technology, while niche cultural or regional subjects are underrepresented. Additionally, the dataset may suffer from temporal bias, lacking significant representation of recent language trends, such as modern internet slang or code-mixed communication styles. These biases could lead the models to prioritize dominant themes, although their overall impact on topic formation appears modest.

Furthermore, the dataset spans a broad range of topics, including movies, technology, and tourism. While this diversity mirrors datasets used in prior studies, model performance may differ on more specialized datasets focused on specific types of short texts, such as reviews for a single product category.

Finally, this study primarily aimed to assess the effectiveness of BERTopic for general Hindi short texts, without targeting specific short text types such as informal conversations or mixed-language content. Future research utilizing datasets with narrowly defined topics or specialized short text variants is recommended to evaluate these models in more targeted contexts.

## Acknowledgments

We sincerely thank the team at College of Engineering Pune Technological University for verifying and endorsing the quality of the results from our topic modeling pipeline. We also extend our heartfelt gratitude to the anonymous reviewers for their insightful feedback and constructive suggestions.